\documentclass{article}
\usepackage{spconf,amsmath,graphicx}
\usepackage{amssymb}
\usepackage{amsfonts}
\usepackage{tabularx}
\usepackage{algorithm}
\usepackage{algpseudocode}
\usepackage{comment}
\usepackage{booktabs}
\usepackage{bm}
\usepackage{graphics}
\usepackage{float}
\usepackage[utf8]{inputenc} 
\usepackage{indentfirst}
\title{Optimizing Trading Strategies in Quantitative Markets using Multi-Agent Reinforcement Learning}
\name{
   \raggedright 
       {
       \fontsize{10}{10}\selectfont
       Hengxi Zhang$^{1\dagger}$\thanks{$^{\dagger}$Contribute equally.}, Zhendong Shi$^{1\dagger}$, Yuanquan Hu$^{1}$, Wenbo Ding$^{1*}$\thanks{$^*$Corresponding author (Email: ding.wenbo@sz.tsinghua.edu.cn).}, Ercan E. Kuruoğlu$^{1}$, Xiao-Ping Zhang$^{1,2}$, IEEE Fellow
       \thanks{This work was supported in part by Shenzhen Science and Technology Program (JCYJ20220530143013030), by Shenzhen Ubiquitous Data EnablingKey Lab (ZDSYS20220527171406015 ), by Guangdong Innovative and Entrepreneurial Research Team Program (2021ZT09L197), and by Tsinghua Shenzhen International Graduate School-Shenzhen Pengrui Young Faculty Program of Shenzhen Pengrui Foundation (No. SZPR2023005).}
       }
}

\address{$^{1}${\fontsize{11}{12}\selectfont 
Shenzhen International Graduate School, Tsinghua University, Shenzhen, China}\\
$^{2}${\fontsize{11}{12}\selectfont 
Department of Electrical, Computer and Biomedical Engineering, Ryerson University, Toronto, Canada}
}

\begin{document}
\maketitle

\begin{abstract}
Quantitative markets are characterized by swift dynamics and abundant uncertainties, making the pursuit of profit-driven stock trading actions inherently challenging. Within this context, Reinforcement Learning (RL) — which operates on a reward-centric mechanism for optimal control — has surfaced as a potentially effective solution to the intricate financial decision-making conundrums presented. This paper delves into the fusion of two established financial trading strategies, namely the constant proportion portfolio insurance (CPPI) and the time-invariant portfolio protection (TIPP), with the multi-agent deep deterministic policy gradient (MADDPG) framework. As a result, we introduce two novel multi-agent RL (MARL) methods: CPPI-MADDPG and TIPP-MADDPG, tailored for probing strategic trading within quantitative markets. To validate these innovations, we implemented them on a diverse selection of 100 real-market shares. Our empirical findings reveal that the CPPI-MADDPG and TIPP-MADDPG strategies consistently outpace their traditional counterparts, affirming their efficacy in the realm of quantitative trading.

\end{abstract}

\vspace{-.3cm}

\section{Introduction}
\vspace{-.2cm}
Compared with traditional trading methods, quantitative trading is widely known for its features of high-frequency, algorithmic, and automated trading, which is difficult to achieve by human beings in such a complex and dynamic stock market~\cite{thakkar2021comprehensive, daniels2003quantitative}.
In the quantitative market, massive noisy signals from stochastic trading behaviors and all kinds of unforeseeable social events make the prediction of the market state grueling~\cite{guo2017quantitative, moskowitz2012time}. 
And human traders can easily be affected by these events as well as their body and psychological conditions, which would make the irrational decisions of trading nearly inevitable~\cite{pham2009continuous, 
fleming2004application}. 
Therefore, different financial individuals and institutes from different research fields have started to explore more effective ways for handling these problems.

Over the past years, with the development of artificial intelligence techniques, reinforcement learning (RL) has emerged as an efficient method for making decisions in dynamic environments with uncertainties~\cite{an2022deep}. 
The principle behind RL is the Markov decision process (MDP).
Through interacting with the environment, the RL agent, i.e. the decision maker, will iteratively update its strategy according to the rewards, which can be treated as guidance toward the expected target and the goal of the RL agent is hence to maximize the total reward~\cite{sutton2018reinforcement}. 
Following the MDP, researchers from financial fields have tried to build their own specifically designed RL architecture to cope with different financial problems.
A deep RL method combined with knowledge distillation was proposed to improve the training reliability in the trading of currency pairs~\cite{tsantekidis2021diversity}.
To investigate the stock portfolio selection problem, a hypergraph-based RL method was designed to learn the policy function of generating appropriate trading actions~\cite{li2022hypergraph}.
Besides, a policy-based RL framework for stock portfolio management was introduced and its performance was also compared with other trading strategies~\cite{zhang2021deep}.

Meanwhile, instead of focusing only on a centralized agent that interacts with the environment, people have begun to find that many scenarios, such as multi-robot control and multi-player games, are more like multi-agent system (MAS), where more than one agent is involved~\cite{zhang2021multi}. 
The framework of multi-agent RL (MARL) is hence established to handle those decision-making and optimal control problems with multiple agents inside a common environment~\cite{busoniu2008comprehensive}. 
Similar to the RL, MARL also concentrates on the issues of sequential decision-making, in which each agent needs to take action with its own strategic brain, which can be naturally designed using neural networks~\cite{mnih2015human}. 
However, the situation becomes more complex since the dynamics of each agent will also be treated as the changes in the environment~\cite{vinyals2019grandmaster}. 
To achieve the management of the portfolio under the continuous changes over the market, a MARL-based system was proposed to maximize the return~\cite{lee2020maps}.
Similarly, an MAS stock market simulator was designed to address the issue of assessment over the market activity and reproduce the market metrics~\cite{lussange2021modelling}.

In this work, we integrate an MARL approach named multi-agent deep deterministic policy gradient (MADDPG) into two prior trading strategies, namely constant proportion portfolio insurance (CPPI) and time-invariant portfolio protection (TIPP) respectively for studying how this novel MAS architecture will behave in quantitative markets. 
The rest of this work is organized as follows. Section II introduces the system model, in which we discuss the MAS model and represent how the MADDPG is specifically established with CPPI and TIPP strategies, respectively. Further, the numerical experiment and results are implemented and analyzed in Section III. Finally, the conclusions are drawn in Section IV.

\vspace{-.3cm}

\section{System model}
In this section, we map the problem of strategic trading in quantitative markets into a MARL task. The principle of MADDPG is introduced first, and then we integrate the MADDPG approach into CPPI and TIPP strategies, respectively.

\vspace{-.3cm}

\subsection{Framework of MARL for strategic trading}

A sequential decision-making problem in the multi-agent scenario can be described as a stochastic game, which can be defined by a set of key elements $\langle N, \mathbb{S},\{\mathbb{A}^{i}\}_{i\in\{1,\cdots, N\}}$, $P,\{R^i\}_{i\in\{1,\cdots, N\}},\gamma \rangle$, where $N$ is the number of agents. At time $t$, under a shared state $S_t\in\mathbb{S}$, each agent takes its action $a^i\in\mathbb{A}^i$ simultaneously. The joint action $\mathbf{a}=\{a^1,\cdots,a^N\}$ leads the environment changes according to the dynamics $P:\mathbb{S}\times\pmb{\mathbb{A}}\rightarrow \Delta(\mathbb{S})$, where $\pmb{\mathbb{A}}:=\mathbb{A}^1\times\cdots\times\mathbb{A}^N$. After that each agent receives its individual reward $r^i$ according to its reward function $R^i:\mathbb{S}\times\pmb{\mathbb{A}}\times\mathbb{S}\rightarrow \mathbb{R}$. $\gamma$ is the discount factor that represents the value of time.

In our model, $N$ agents employ diverse strategies to trade in quantitative markets. They aim to optimize their returns while ensuring portfolio diversity, thereby distributing risks among all agents. At every step, agents observe a shared state $\bm{s}=\{s_1,\cdots,s_N\}$, where the individual state $s=[p, h, b]$ is a vector that includes $D$ kinds of stock price $p \in \mathbb{R}_{+}^{D}$, share $h \in \mathbb{Z}_{+}^{D} $, and the remaining balance $b \in \mathbb{R}_{+}^{D}$. Then each agent determines an action $a$, which is a vector representing the weight of each stock in the portfolio, according to its policy. After taking the joint action, the number of shares of each agent is modified and their portfolios are updated. And each agent receives a reward based on the asset value change. 

\subsection{Strategies of CPPI and TIPP} %
\begin{figure}[t!]
\centering
\includegraphics[width=\linewidth]{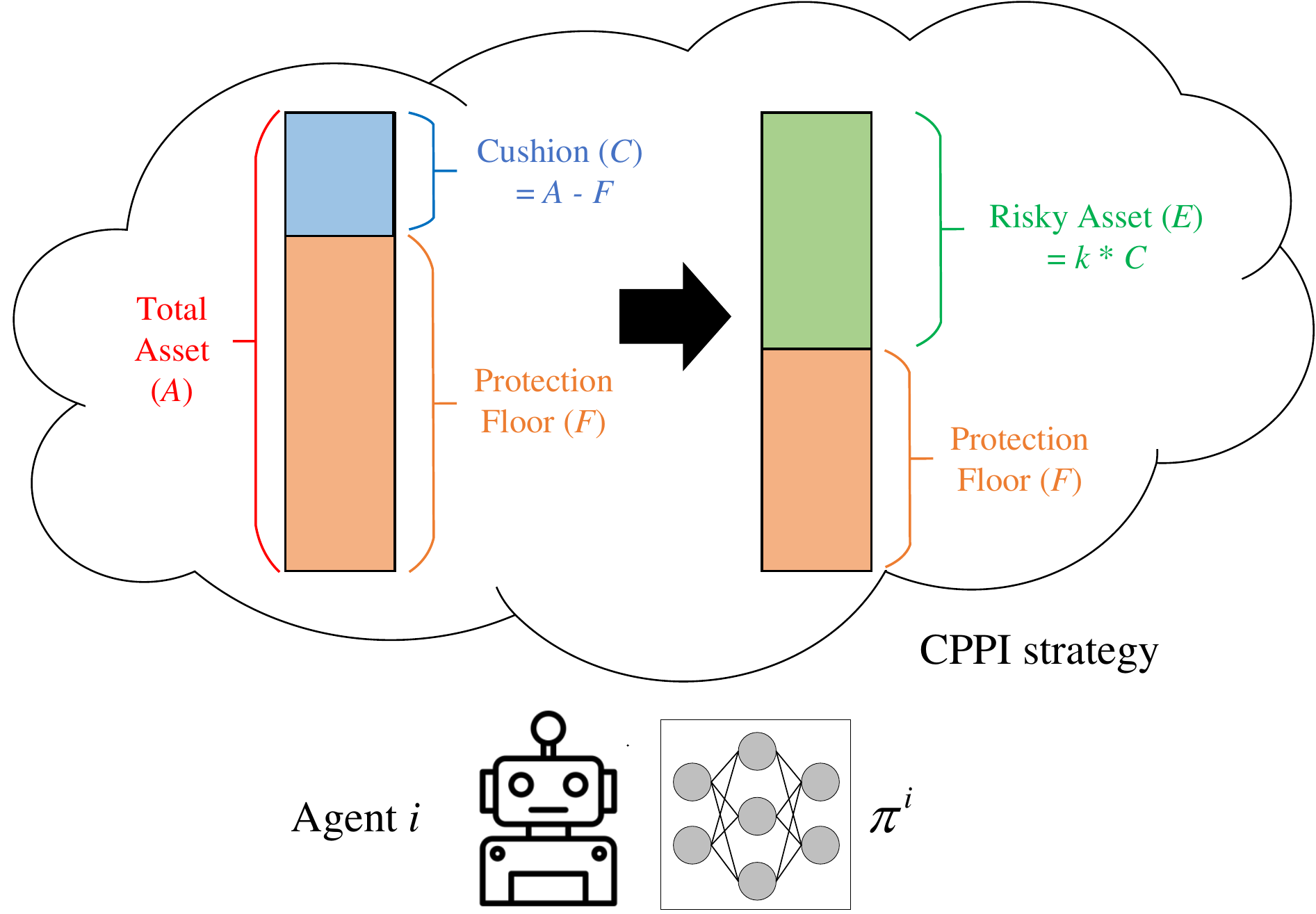}
\caption{The principle of CPPI strategy for agent $i$.}
\label{CPPI}
\end{figure}

Constant Proportion Portfolio Insurance (CPPI) is a type of portfolio insurance in which the investor sets a floor based on their asset, then structures asset allocation around the trading decision~\cite{balder2009effectiveness}. 
As shown in Fig.~\ref{CPPI}, the total asset $A$ is separated into two parts, the protection floor $F$ and the cushion $C$, in which the floor $F$ is the minimum guarantee used for protecting the basis of the total asset and the multiple cushions $k*C$ is supposed to be used as the risky asset $E$,
\begin{equation}
    E = k * C = k * (A - F),
    \vspace{-.2cm}
\end{equation}
where the risk factor $k$ indicates the measurement of the risk and a higher value denotes a more aggressive trading strategy. As a comparison, Time-Invariant Portfolio Protection Strategy (TIPP) is a variation of CPPI, where the protection floor $F_t$ at time step $t$ is not a fixed value and changes over time according to some percentage of the total asset $A$ and the previous floor $F_{t-1}$,
\begin{equation}
\begin{split}
    F_t = max\{ \phi A_t, F_{t-1}\},\\
    E_t = k (A_t-F_t),
\end{split}
\vspace{-.2cm}
\end{equation}
where $\phi$ is the floor percentage. As the total asset $A$ increases, the amount of guarantee will accordingly rise. While the guarantee remains unchanged if the portfolio reduces.

\subsection{Multi-Agent Deep Deterministic Policy Gradient with Insurance Strategy}

\begin{figure}[t!]
\vspace{-1cm}
\centering
\includegraphics[width=.95\linewidth]{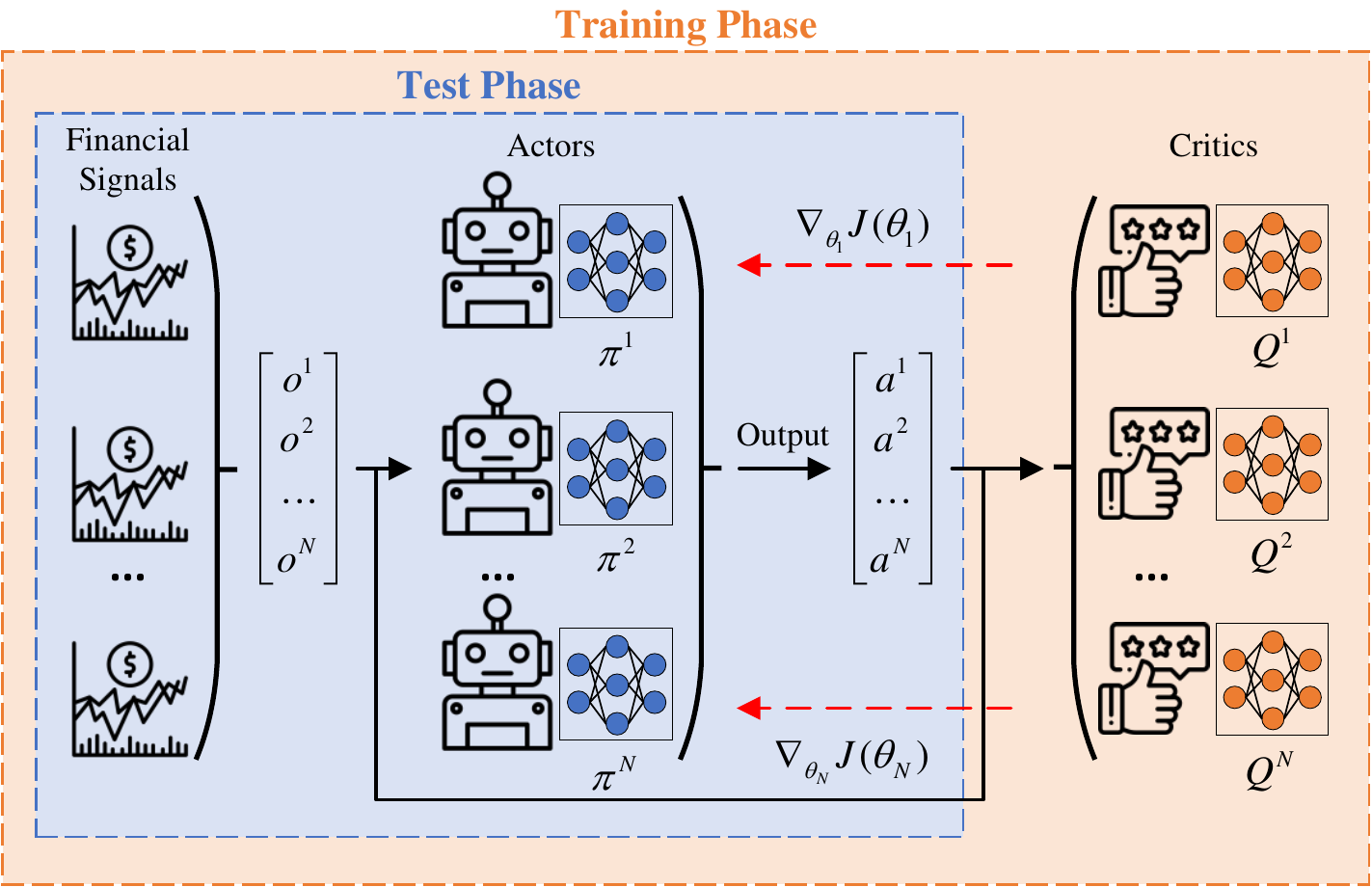}
\caption{A schematic of MADDPG in the quantitative market environment.}
\label{MADDPG_frame}
\end{figure}

We adopt the MADDPG~\cite{lowe2017multi} to train our agents. This approach is specifically established for the implementation in the scenario of quantitative trading, presented as Fig.~\ref{MADDPG_frame}. MADDPG is a multi-agent version of actor-critic method considering a continuous action set with a deterministic policy, where each agent has an actor network $\pi_\theta$ parameterized by $\theta$ and a critic network $Q_\phi$ parameterized by $\phi$. Each agent learns its optimal policy by updating the parameters of its policy networks to directly maximize the objective function, i.e., the cumulative discounted return,
$J(\theta_i)=\mathbb{E}_{s\sim P,\bm{a}\sim\bm{\pi}_\theta}[\sum_{t\geq 0}\gamma^t R_t^i]$ and the direction to take steps by agent $i$ can be presented as the gradient of the cumulative discounted return, shown as,
\vspace{-.5cm}

\begin{equation}\label{actor loss}
    \begin{split}
    \nabla_{\theta_i}J(\theta_i)=\mathbb{E}_{s\sim\mathcal{D}}\bigg[\nabla_{\theta_i}\log \pi_{\theta_i}(a_i|s)\cdot\\
    \nabla_{a_i}Q_{\phi_i}(s,a_1,\cdots,a_N)|_{a_i=\pi_{\theta_i}(s)}\bigg],
    \end{split}
\end{equation}
where $\mathcal{D}$ is the experience replay buffer containing tuples $(s,s',a_{1},\cdots,a_{N},r_{1},\cdots,r_{N})$ that are stored throughout training. To combine the policy with the insurance strategy, we use CPPI or TIPP to adjust the output actions instead of using raw outputs as the actions. The feasibility of such adjusting also benefits from the off-policy nature of MADDPG.

The centralized critic networks are updated by approximating the true action-value function using temporal-difference learning. Further, in order to avoid all agents moving towards the same strategy, we set the correlation between agents as part of the loss function to achieve the purpose of portfolio selection. The loss function for agent $i$ can be expressed as,
\begin{equation}\label{critic loss}
\begin{split}
     \mathcal{L}
    (\phi_i)= \lambda \mathbb{E}_{s,\bm{a},\bm{r},s'}\bigg[(Q_{\phi_i}(s,a_1,\cdots,a_N)-y)^2\bigg] &\\+ (1-\lambda) \sum_{i=1,i \neq j}^{K}Corr(a_i,a_j)^2, \\
    y=r_i+\gamma Q_{\phi'_{i}}(s',a'_{1},\cdots,a'_{N})|_{a'_{j}=\pi_{\theta'_{i}}(s)},
\end{split}
\end{equation}
where $Q_{\phi'_{i}}$ and $\pi_{\theta'_{i}}$ are target networks with delayed parameters $\phi'_{i}$ and $\theta'_{i}$, $a_i$ is the action vector showing the positional confidence vector of agent $i$ under the restriction of CPPI or TIPP, and $\lambda$ is the hyperparameter controlling the equilibrium. The pseudocode of the proposed algorithm is shown in Alg.~\ref{alg:1}. Notice that the rule-based policy $\Phi$ here can be either CIPP or TIPP, which outputs the constraints of the actions under the insurance strategy.

\section{Experiments}
\vspace{-.2cm}
We conduct the experiments on a real-world stock trading environment provided by FinRL~\cite{liu2021finrl}. We test the performance of our proposed CPPI-MADDPG and TIPP-MADDPG with MADDPG, MADQN, and the Universal Portfolio (UP). The experimental results show that combining MADDPG with insurance strategy provides a substantial advantage in the realm of quantitative trading. 

\subsection{Dataset and Settings}
In our experiment, we selected 100 stocks listed on the Shenzhen Stock Exchange, sourced via Tushare. These stocks have codes ranging from 000010.SZ to 300813.SZ. When combined with cash as a risk-free asset, the potential array of investment products expands significantly. For training, we utilized data spanning from January 1st, 2018 to December 31st, 2020. The testing set comprises data from January 1st, 2021 to December 31st, 2021. To align our analysis with real-world market conditions, constraints were applied to the data, including non-negative balance maintenance and the incorporation of transaction costs. We initialized with a set cash amount, aiming to maximize profits using the aforementioned trading strategies.

\begin{algorithm}[t!]
\caption{MADDPG with insurance strategy}\label{alg:1}
\begin{algorithmic}[1]
\State Initialize the rule-based policy $\Phi$ according to the insurance strategy.
\State Initialize $Q_{\phi_i}$, $\pi_{\theta_i}$, $Q_{\phi'_i}$, $\pi_{\theta'_i}$ for all $i\in\{1,\cdots,N\}$.
\While {training not finished}
    \State Initialize initial state $s$ and a random process $\mathcal{N}$ for\newline
    \hspace*{1.25em} action exploration. 
    \For {each episode}
        \State For each agent select action $a_i=\pi_{\theta_i}(s)+\mathcal{N}_t$. 
        \If {$a_i\notin \Phi(s)$}
            \State adjust $a_i$ with $\Phi(s)$.
        \EndIf
        \State Execute joint action $\bm{a}$ and observe reward $\bm{r}$ and \newline
        \hspace*{2.75em} next state $s'$.
        \State Store experience $\langle s,\bm{a},\bm{r},s' \rangle$ in replay buffer $\mathcal{D}$.
        \State Sample a minibatch of $K$ experiences from $\mathcal{D}$.
        \For {each agent}
            \State Update the critic $Q_{\phi_i}$ by minimizing Eq.(\ref{critic loss}).
            \State Update the actor $\pi_{\theta_i}$ using Eq.(\ref{actor loss}).
        \EndFor
        \State Update target networks for each agent:
        \begin{equation*}
            \begin{split}
                \phi'_i\leftarrow \tau \phi_i+(1-\tau)\phi'_i,\\
                \theta'_i\leftarrow \tau \theta_i+(1-\tau)\theta'_i.
                \vspace{-.2cm}
            \end{split}
        \end{equation*}
    \EndFor
\EndWhile
\end{algorithmic}
\end{algorithm}
\vspace{-.5cm}

\subsection{Results}

\begin{figure*}[ht]
    \centering
    \includegraphics[width=.85\linewidth]{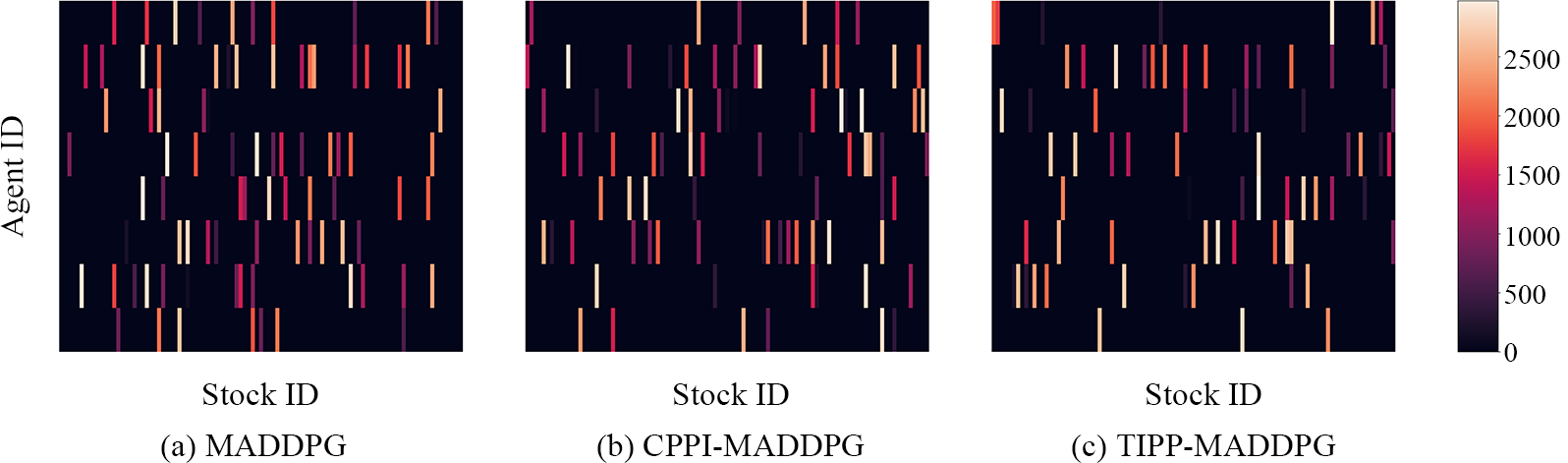}
    \caption{The asset allocations with MADDPG, CPPI-MADDPG, and TIPP-MADDPG strategies .}
    \vspace{-.5cm}
    \label{fig:theo}
\end{figure*}

\begin{figure}[ht]
\centering
\includegraphics[width=\linewidth]{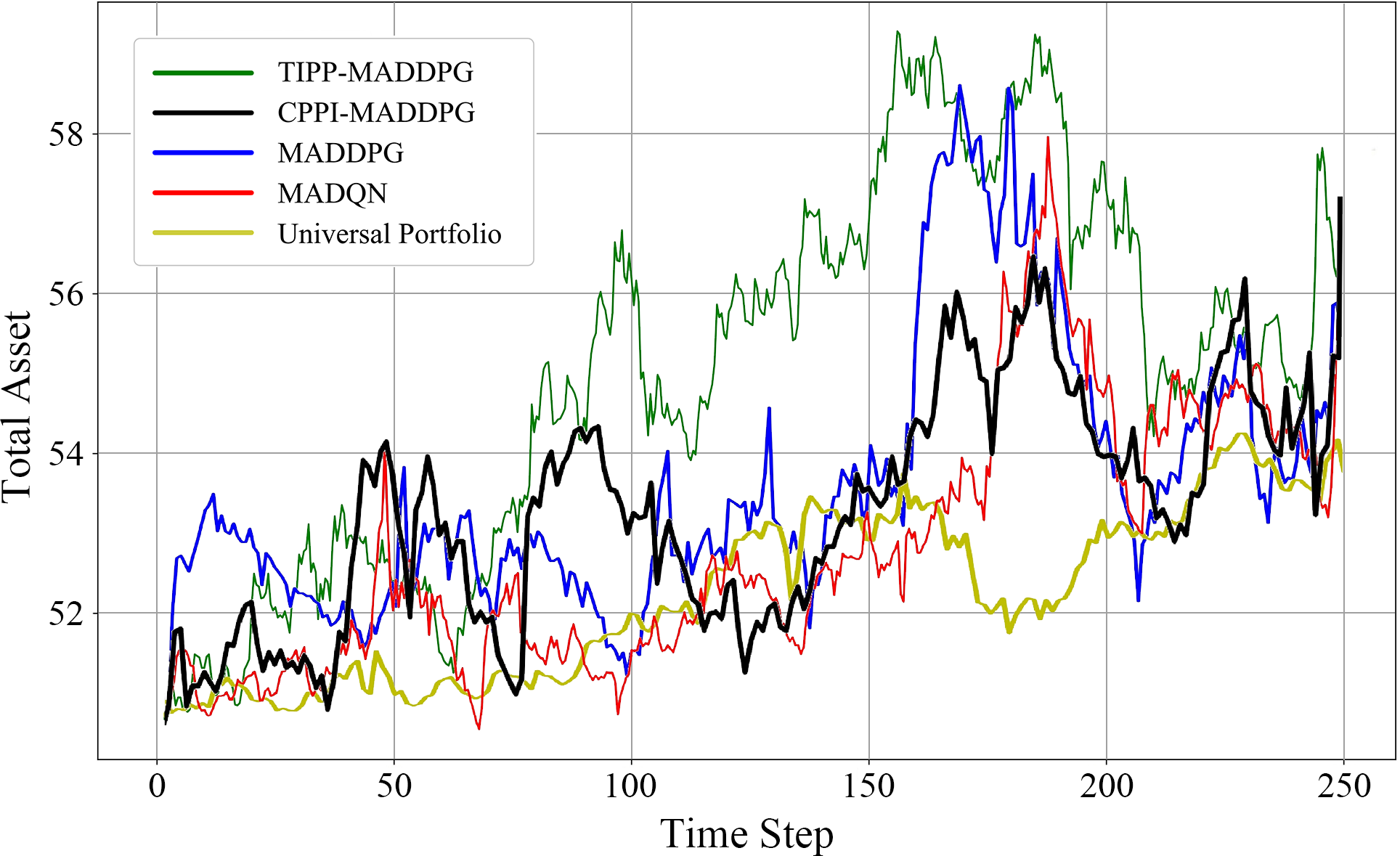}\caption{The performances of portfolios with different strategies (The metrics of Total Asset and Time Step: $10^3$ RMB and Day).}
\label{reward}
\vspace{-.4cm}
\end{figure}

The changes in the total assets with different trading strategies over time are present in Fig.~\ref{reward}, where only the trading days are leveraged for the algorithm implementation. In the first 80 trading days, the profits gained by all strategies except for Universal Portfolios (UP), which has increased in a stable but slow manner, are nearly the same. Afterward, TIPP-MADDPG has boomed during the 80th to 160th trading days compared with CPPI-MADDPG method. One of the significant risks with CPPI is the "gap risk". If the risky asset falls dramatically in value in a very short period, the portfolio might not be rebalanced in time to prevent it from breaching the floor value. TIPP, with its time-based adjustment, does not have this risk to the same extent.

We then compare the efficacy of our proposed trading strategies against UP, MADQN, and MADDPG using metrics such as Annual Return (AR), Sharpe Ratio (SR), and Maximum Drawdown (MaxD, which represents the maximum portfolio value drop from peak to trough). For MADQN, we use discretized action space as in MADDPG. The UP is a standard portfolio method, which optimizes based on the correlation of diverse stock returns. The results for AR, SR, and MaxD are detailed in Table~\ref{tab:data}. 

From the numerical results, we can see that UP struggles with real-time data, underperforming in our test set. MADQN's limitation is its restricted exploration capacity, making it less adept in a stock market rife with uncertainties. Both CPPI-MADDPG and TIPP-MADDPG can revert to the baseline MADDPG strategy under certain parameters, allowing them to be tailored to individual investor risk preferences. For profitability, the TIPP strategy mandates a greater capital guarantee than CPPI, leading to a more assertive portfolio approach. This results in TIPP excelling in annual return, whereas CPPI outshines in SR and MaxD metrics.

To further study how each agent has made a series of trading decisions over time in the test phase, we visualize the general trading behavior for each agent on 100 shares with MADDPG, CPPI-MADDPG, and TIPP-MADDPG, respectively. 
As shown in Fig.~\ref{fig:theo}, the heatmap presents how the agents with different strategies choose to allocate the asset. The agents with MADDPG prefer the relatively uniform allocation while the assets allocated by those with CPPI-MADDPG and TIPP-MADDPG are more sparse.

\begin{table}[t]
    \normalsize
    \caption{Comparison of Different Strategies}
    \label{tab:data}
    \centering
    \resizebox{0.88\linewidth}{!}{
        \centering
        \begin{tabular}{c c c c}
          \toprule
          Strategy & AR & SR & MaxD \\
          \midrule 
          UP & $3.36\%$ & $1.03$ & $3.86\%$\\
          MADQN & $6.47\%$ & $1.71$ & $9.43\%$\\
          MADDPG & $8.22\%$ & $1.99$ & $\mathbf{12.26\%}$\\
          CPPI-MADDPG & $7.76\%$ & $\mathbf{2.18}$ & $6.60\%$\\
          TIPP-MADDPG & $\mathbf{9.68\%}$ & $2.09$ & $9.02\%$\\
          \bottomrule
        \end{tabular}
    }
\end{table}

\vspace{-.2cm}
\section{Conclusion}
In this study, we introduced two augmented MADDPG algorithms: CPPI-MADDPG and TIPP-MADDPG. Our goal was to elucidate the advantages these financial trading strategies offer when integrated with MARL for quantitative trading. We subjected both methods to rigorous testing using genuine financial data from the Shenzhen Stock Exchange. The empirical outcomes underscore the efficacy of our approaches. We believe these findings highlight the potential for future deployments of our methods in quantitative markets. Given the significant outcomes and the evolving nature of quantitative markets, we are optimistic about the broader applications and adaptations of our methodologies.


\bibliographystyle{IEEEbib}
\bibliography{strings}

\end{document}